\documentclass[conference]{IEEEtran}
\usepackage{cite}
\usepackage{amsmath,amssymb,amsfonts,booktabs}
\usepackage{algorithmic}
\usepackage{graphicx}
\usepackage{textcomp}
\usepackage{xcolor}
\def\BibTeX{{\rm B\kern-.05em{\sc i\kern-.025em b}\kern-.08em
    T\kern-.1667em\lower.7ex\hbox{E}\kern-.125emX}}
\begin{document}

\title{Deep Probabilistic Modelling of Price Movements for High-Frequency Trading}

\author{\IEEEauthorblockN{Ye-Sheen Lim}
\IEEEauthorblockA{\textit{Department of Computer Science} \\
\textit{University College London}\\
London, UK \\
yesheenlim@gmail.com}
\and
\IEEEauthorblockN{Denise Gorse}
\IEEEauthorblockA{\textit{Department of Computer Science} \\
\textit{University College London}\\
London, UK \\
d.gorse@cs.ucl.ac.uk}
}


\maketitle

\begin{abstract}
In this paper we propose a deep recurrent architecture for the probabilistic modelling of high-frequency market prices, important for the risk management of automated trading systems. Our proposed architecture incorporates probabilistic mixture models into deep recurrent neural networks. The resulting deep mixture models simultaneously address several practical challenges important in the development of automated high-frequency trading strategies that were previously neglected in the literature: 1) probabilistic forecasting of the price movements; 2) single objective prediction of both the direction and size of the price movements. We train our models on high-frequency Bitcoin market data and evaluate them against benchmark models obtained from the literature. We show that our model outperforms the benchmark models in both a metric-based test and in a simulated trading scenario.
\end{abstract}

\section{Introduction}

The aim of this paper is to develop a deep neural network architecture for producing forecasts that is suitable for use in an automated trading strategy (ATS) pipeline to reduce trading risk. When designing and developing an ATS in practice, \emph{risk management} is arguably a more crucial part of the pipeline than forecasting. In a complex dynamical system such as a financial market, it is sensible to expect that even the best trading algorithms, that incorporates state-of-the-art models and are driven by the most expensive and exclusive datasets, will suffer losses at some point. If risk is improperly managed, for example by holding large positions in assets with high variance, the trader may rapidly lose all their capital.

When the forecasting component of an ATS has discovered a trading opportunity, the information is passed on to the risk management component to quantitatively determine (based on current trade positions and capital) if the trading opportunity should be taken and, if so, how much capital should be assigned to this opportunity. In order to make these decisions, uncertainties surrounding the trading opportunities need to be known. These uncertainties can either be extrinsically obtained through the forecast errors of historical backtesting, or intrinsically obtained from \emph{probabilistic forecasts}. 

Uncertainties obtained from backtesting are useful to manage risks in trading strategies such as the long-short strategy \cite{goumatianos2013stock}, which theoretically guarantees no losses. However, probabilistic forecasts can directly be fed into a much greater variety of proven and well-established industry standards such as the computation of Value at Risk (VaR) \cite{jorion2000value} and the Kelly Criterion\cite{thorp2011kelly}. VaR, which computes the potential loss of a trading opportunity, is directly known if the trading opportunity is described in the form of a probability distribution. Probabilistic forecasts also allows us to compute of the Kelly Criterion, which suggests the optimal bet sizes for a given trading opportunity require knowledge of the win or lose probabilities, as well as the expected values of the win and loss.

In this paper we propose a novel architecture that combines deep learning and statistical modelling to address the main challenge of the probabilistic forecasting of high-frequency price movements. The nature of the architecture also allows us to address the secondary challenge of determining the size of price movements. In finance, it is more common to treat forecasting as a classification problem rather than a regression problem. Knowing if the price will go up or down is vital as it gives us the all-important information as to which position (to buy or to sell) to take on a trading opportunity. However, along with this directional information, knowing also by how much the price will go up or down (i.e. the size of the movements) will give the trader a much more complete picture of risk and therefore enable better decision making. When computing the Kelly Criterion, for instance, we would require probability distributions of the size of the price movements.

This paper is organised as follows. In Section \ref{section:related_work} we present a review of current literature related to forecasting for high-frequency trading and to deep probabilistic forecasting. Then in Section \ref{section:method} we describe our proposed architecture. Our data collection, experimental set-up, model implementation, and test and benchmarking procedures are described in Section \ref{section:experimental_setup}. The next section \ref{section:results} presents the results of the experiments. Then, in Section \ref{section:trading_scenario}, we present a simulated trading scenario using real-world data in which we test the practical usage of model and its benchmarks. Finally we conclude with a summary and discussion of future work in Section \ref{section:discussion}.

\section{Related Work}
\label{section:related_work}

In the domain of high-frequency trading (HFT), the most common method for producing probabilistic forecasts is by the stochastic modelling of price movements, an extensively researched topic in quantitative finance. \cite{bacry2014hawkes,toke2011introduction,cont2010stochastic,jayasekare2016modeling} These approaches are however not extensively data-driven and rely on fitting closed-form theory-driven stochastic models, leading to major drawbacks such as sensitivity to regime shifts, intractability and lack of generalisation power. Due to these drawbacks, trading strategies driven by these models rely heavily on risk management to limit potential losses.

These kinds of drawbacks can be overcome with data-driven approaches using machine learning models. Deep learning models in particular have exhibited state-of-the-art results in predicting high-frequency stock price movements \cite{sirignano2019universal,tsantekidis2017using,sirignano2019deep}. However, the needs of risk management have been mostly overlooked in the existing literature. These works are focused on producing deterministic forecasts, meaning that for risk management the trader has to rely on the less useful measure of uncertainty that is extrinsically obtained from historical backtesting. For instance, in absence of probabilistic forecasts a method of using the backtesting false positives and false negatives rates to compute profit risk has been proposed \cite{dixon2018sequence}. Furthermore, all previous deep learning papers predict only the \emph{directional} price movements and do not provide any output on the size of the price movements. The only exception is \cite{sirignano2019deep}), whose output can be used to compute the size of the price movements.

Probabilistic forecasts can be obtained from deep learning models either by exploiting dropout to compute uncertainty \cite{gal2016dropout}, or by adopting a probabilistic network architecture. For the application presented in this paper, the probabilistic architecture approach allows for the specification of application-suitable probabilistic models and likelihood functions, and also reduces the time it takes to produce the probabilistic forecast, which is a crucial process in HFT. Such probabilistic architectures have been shown to be successful in the domains of e-commerce \cite{wen2017multi, salinas2019deepar} and NLP \cite{mei2017neural}. The novelty in our proposed architecture compared to existing work is in the way the output is specified to produce forecasts suitable for use in automated HFT pipelines.

\section{Method}
\label{section:method}

\subsection{Problem Formulation}

HFT data is characterized by the \emph{order flow}, which is the microsecond stream of events arriving into an exchange. Each event is essentially an action taken by a market participant, such as placing and cancelling orders. The order flow is the underpinning mechanism behind all the changes we see on the price charts one would find on Google Finance, for instance. Readers are directed to \cite{gould2013limit} for a comprehensive introduction to this concept as its details are outside the scope of this paper. It should be noted that we change much of the jargon used in \cite{gould2013limit} to more intuitive terms for readers without a background in quantitative finance.

Let us denote an order flow of length $m$ as an irregularly spaced sequence $\mathbf{x}_{i,:m}$. Since events can arrive, and price change can occur, at any time, irregularly spaced sequences are an important characteristic of HFT data. Given $\mathbf{x}_{i,:m}$, we want to predict the price change after the last event of $\mathbf{x}_{i,:m}$. The word "after" is loosely defined here such that the price change could be caused by the next event, or be the price change $\tau$ seconds after the last event in the stream, and so on. Denoting the price change as $y_{i,m}$, our main goal is to then model the conditional distribution of $y_i$

\begin{equation}
P(y_{i,m} | y_{i,:m-1}, \mathbf{x}_{i,:m}, \mathbf{x}^{(s)}_i)
\label{eq:conditional_distribution}
\end{equation}

\noindent
where $x_i$ are static covariates, $\mathbf{x}_{i,:m}$ are the non-autoregressive temporal covariates and $y_{i,:m-1}$ are the autoregressive temporal covariates. Note here that since the temporal covariates $\mathbf{x}_{i,:m}$ have irregularly spaced intervals, each of the auto-regressive covariates $y_{i,:m-1}$ may be computed relative to the timestamp of the temporal covariates. For example, if $y_{i,m}$ is the price change $\tau$ seconds after $x_{i,m}$, then $y_{i,m-1}$ is the price change $\tau$ seconds after $x_{i,m-1}$.

High-frequency price movements are inherently discrete since the rules of any given stock exchange typically define the \emph{tick}, which is the minimum amount by which prices can change \cite{gould2013limit}. Therefore, we have $y_{i,m} \in \mathbb{Z}$ and can formulate our problem as one of modelling count data (i.e. how many ticks the price has moved up or down by).

\subsection{Proposed Architecture}

\begin{figure}[ht!]
\centerline{\includegraphics[width=0.6\columnwidth]{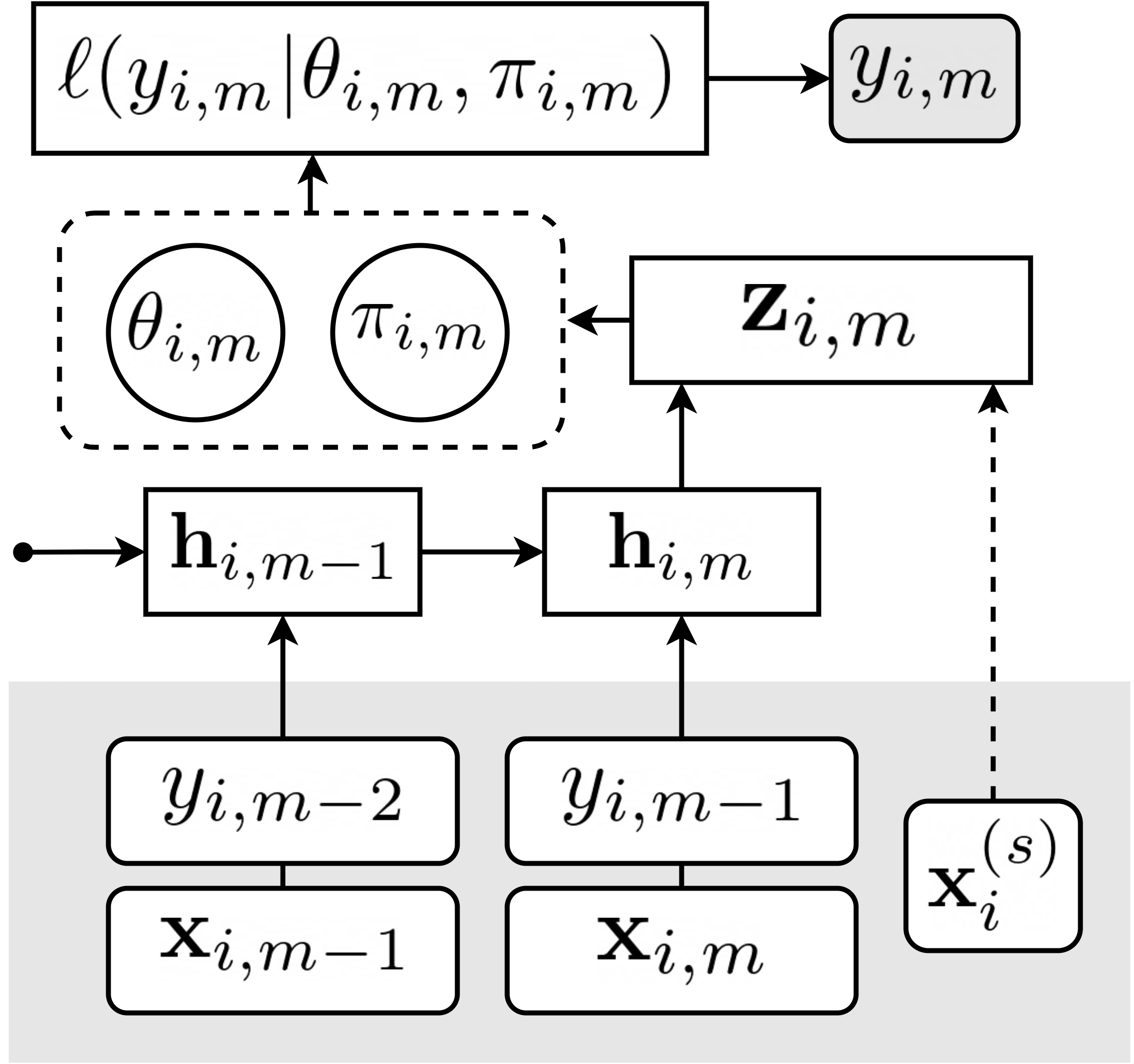}}
\caption{Proposed architecture}
\label{fig:architecture}
\end{figure}

Our proposed architecture for modelling the conditional distribution in Equation \ref{eq:conditional_distribution} is summarised in Figure \ref{fig:architecture}.

Let $h(.)$ be a function implementing a deep recurrent neural network (RNN). We first learn abstract representations of the temporal covariates $\mathbf{x}_{i,:m}$ and $y_{i,:m-1}$ using an $L$-layer stacked recurrent neural network (RNN). The output at each layer $l$ can be described as follows

\begin{equation}
  \mathbf{h}^l_{i,m} =
  \begin{cases}
    h( \mathbf{h}^{l-1}_{i,m}, \mathbf{h}^{l}_{i,m-1}, \Theta^l ) & \text{if $1 < l \leq L$} \\
    h( y_{i,:m-1}, \mathbf{x}_{i,m}, \mathbf{h}^{l}_{i,m-1}, \Theta^l )  & \text{if $l=1$} \\
  \end{cases}
\label{eq:rnn}
\end{equation}

\noindent
where $\Theta_l$ are the neural network parameters associated with layer $l$. Any individual covariates $x^j_{i,m} \in \mathbf{x}_{i,m}$ that are non-ordinal and categorical are embedded beforehand into a multidimensional continuous space before feeding them into the inputs of the RNNs. This embedding, denoted $\tilde{x}^j_{i,m}$, is implemented as follows

\begin{equation}
  \tilde{x}^j_{i,m}= g\left( {\mathbf{W}^j}^\intercal o(x^j_{i,m}) +\mathbf{b}^j \right)
  \label{eq:embedding}
\end{equation}

\noindent
where $o(.)$ is a function mapping the categorical features to one-hot vectors, $g(.)$ is some non-linear activation function, and $\mathbf{W}_q$ and $\mathbf{b}_q$ are parameters to be fitted.

How the static covariates $\mathbf{x}^{(s)}_i$ are treated depends on the cardinality. If the cardinality is small, then we append them at each point in the sequence and repeat them over the whole sequence. On the other hand, if the cardinality is large then the previous method would be inefficient and we would feed $x_i$ into a dense layer implemented similarly to Equation \ref{eq:embedding} but without the one-hot mapping function.

Whether or not it is concatenated with $h^{(s)}_i$, $\mathbf{h}^L_{i,m}$ is then fed into a D-layer fully connected neural network for a final abstraction step. The output at each layer $d$ of this step are computed as follows

\begin{equation}
  \mathbf{z}^d_{i,m} =
  \begin{cases}
    g\left( {\mathbf{W}^d}^\intercal \mathbf{h}^L_{i,m} +\mathbf{b}^d \right) & \text{if $d=1$} \\
    g\left( {\mathbf{W}^d}^\intercal \mathbf{z}^{d-1}_{i,m} +\mathbf{b}^d \right) & \text{if $1 < d \leq D$} \\
  \end{cases}
\label{eq:static_dense}
\end{equation}

\noindent
where $\mathbf{h}^L_{i,m})$, $g(.)$ is some non-linear activation function, and $W_l$ and $b_l$ are parameters to be fitted. 

To obtain a probabilistic forecast suitable for use in an automated high-frequency trading strategy, we propose the novel application of mixture models \cite{mclachlan2004finite} in the output of the architecture for describing the price movements $y_{i,m}$. The mixture probabilities $\mathbf{\pi}_{i,m}$ and the parameters $\mathbf{\theta}_{i,m}$ of the probability distributions are defined as functions of the dense layer output $\mathbf{z}^D_{i,m}$. The model is then fitted by minimising the negative log-likelihood of $y_{i,m}$ given $\mathbf{\pi}_{i,m}$ and $\mathbf{\theta}_{i,m}$.

The choice of the type of mixture models, and consequently the likelihood function, depends on the statistical properties of the data. For our experiments, we consider three choices for the modelling of mixtures of count data: 1) the Poisson mixture model, 2) the Negative Binomial mixture model, 3) the Zero-Truncated Poisson mixture model.

The most common approach to modelling count data, assuming equal mean and variance, is to use the standard Poisson distribution \cite{coxe2009analysis}, and so we will begin by considering this model. Let us denote the mixture probability for component $k$ as $\pi^k_{i,m}$, the Poisson rate parameter for component $k$ as $\lambda^k_{i,m}$, and the number of mixture components as $K$. Also, let $k=1$ correspond to downward price movement and $k=2$ to upward price movement. Given the dense layer output $\mathbf{z}^D_{i,m}$, we define the log-likelihood $\ell_P$ of the 2-component Poisson mixture model used in our experiments as follows, omitting all $i,m$ labels for readability:

\begin{equation}
  \pi^k = \frac{e^{ {\mathbf{W}^{(\pi^k)}}^\intercal \mathbf{z}^D +\mathbf{b}^{(\pi^k)} }}{\sum^{K} e^{ {\mathbf{W}^{(\pi_k)}}^\intercal \mathbf{z}^D +\mathbf{b}^{(\pi_k)} } }
  \label{eq:poisson_mixture}
\end{equation}

\begin{equation}
  \lambda^k =  log( 1 + e^{ {\mathbf{W}^{(\lambda^k)}}^\intercal \mathbf{z}^D +\mathbf{b}^{(\lambda^k)} } )
  \label{eq:poisson_rate}
\end{equation}

\begin{equation}
  \ell_P (y | \mathbf{\pi}, \mathbf{\lambda}) = log \left( \sum^K_k \pi^k \frac{ {\lambda^k}^{|y|} e^{-\lambda^k}}{|y|!} \mathbb{I}_{p(y)=k} \right)
  \label{eq:poisson_likelihood}
\end{equation}

\noindent
In the above equations, $W^{(.)}$ and $b^{(.)}$ are neural network parameters to be fitted, $p(.)$ is a function mapping the sign of $y_{i,m}$ to $k$, and $\mathbb{I}_{(.)}$ is an indicator function for the given statement.

However, much real-world data exhibits \emph{overdispersion}, where the variance is higher than the mean, causing the Poisson distributions to be unsuitable since they do not specify a variance parameter. We should therefore consider an alternative approach in this case, which is to use the standard Negative Binomial distribution \cite{coxe2009analysis}. Using a similar notation style as in the Poisson mixture model definitions, and letting $\mu^k_{i,m}$ and $\alpha^k_{i,m}$ be the mean and shape parameters respectively, we define the log-likelihood $\ell_{NB}$ of the 2-component Negative Binomial mixture model used in our experiment as follows (with $i,m$ once again omitted for readability):

\begin{equation}
  \mu^k =  log( 1 + e^{ {\mathbf{W}^{(\mu^k)}}^\intercal \mathbf{z}^D +\mathbf{b}^{(\mu^k)} } )
  \label{eq:nb_mu}
\end{equation}

\begin{equation}
 \alpha^k =  log( 1 + e^{ {\mathbf{W}^{(\alpha^k)}}^\intercal \mathbf{z}^D +\mathbf{b}^{(\alpha^k)} } )
  \label{eq:nb_alpha}
\end{equation}

\begin{equation}
  \ell_{NB} (y | \mathbf{\pi}, \mathbf{\mu}, \mathbf{\alpha}) = log \left( \sum^K_k \pi_k \eta_1 \eta_2 \mathbb{I}_{p(y)=k} \right)
  \label{eq:nb_mixture_likelihood}
\end{equation}

\begin{equation}
  \eta_1 = \frac{\Gamma(|y|+\frac{1}{\alpha^k})}{\Gamma(|y|+1)\Gamma(\frac{1}{\alpha^k})}
  \label{eq:nb_likelihood_1}
\end{equation}

\begin{equation}
  \eta_2 = {\left( \frac{1}{1+\alpha^k\mu^k} \right)}^{\frac{1}{\alpha^k}} {\left( \frac{\alpha^k\mu^k}{1+\alpha^k\mu^k} \right)}^{|y|}
  \label{eq:nb_likelihood_2}
\end{equation}

\noindent
In the above equations, $W^{(.)}$, $b^{(.)}$ and $p(.)$ are defined as in the Poisson mixture model, and in addition the method of computation for the mixture probabilities $\pi_k$ is not defined here since it is exactly the same as in Equation \ref{eq:poisson_mixture}.

While both the Poison and Negative Binomial mixture models allow for cases when there is no price change, they do not explicitly model $y_{i,m}=0$. We therefore also experiment with a 3-component Zero-Truncated Poisson \cite{ridout1998models} mixture model to explicitly model cases when there is no price change. With mixture component $k=3$ representing $y_{i,m}=0$, we define the likelihood $\ell_{ZP}$ of this mixture model as follows (omitting $i,m$ once again)

\begin{equation}
  \ell_{ZP} (y | \mathbf{\pi}, \mathbf{\lambda}) =  log \left( (\pi^3\mathbb{I}_{p(y)=3} \sum^2_k \pi^k \frac{ {\lambda^k}^{|y|} }{(e^{\lambda^k}-1)|y|!} \mathbb{I}_{p(y)=k} \right)
  \label{eq:zero_poisson_likelihood}
\end{equation}

\noindent
where the mixture probabilities $\pi^k$ and rate parameters $\lambda^k$ are as defined in Equations \ref{eq:poisson_mixture} and \ref{eq:poisson_rate}.

\section{Experimental Set-Up}
\label{section:experimental_setup}

\subsection{Data}

HFT data from stock exchanges are usually very expensive or difficult to obtain for the typical market participant. On the other hand, digital currency exchanges provide participants with APIs that allow, through a websocket feed, access to the same type of high-frequency data as used for HFT in stock exchanges. The data for our experiment is thus obtained from Coinbase, a digital currency exchange. Using the Coinbase API, we a log real-time message stream containing trades and orders updates for \emph{currency pairs} BTC-USD and BTC-EUR between 4 Nov 2017 to 22 Dec 2017.

Since these are raw JSON messages, data for training and testing the model cannot be obtained directly from the message stream. For each of the currency pair, we implement an emulator which sequentially processes the JSON messages to update the state of the exchange over time. From this, we are able to compute different measures to obtain the covariates and target variables necessary for building the training and test sets. We combine the datasets obtained from BTC-USD and BTC-EUR to jointly train the model on both currency pairs. Since we are predicting numbers of ticks, there is no issue with combining datasets of different currency pairs, and in this way we are able to bridge different but very closely related statistical behaviours and obtain a richer dataset.

The combined BTC-USD and BTC-EUR dataset is then split into training, validation and test sets by dates, to avoid any look-ahead bias. Datapoints timestamped between 6 Nov 2017 and 15 Nov 2017 are taken as the training set. Data prior to 6 Nov 2017 are not available since they were needed to "warm up" the emulator. Cross-validation and early-stopping checks are performed on datapoints taken from 16 Nov 2017 to 18 Nov 2017. The rest of the datapoints (from 19 Nov 2017 to 22 Dec 2017) are kept strictly for testing and are never seen by the model aside from this final testing phase. There are about 2 million, 0.7 million and 6 million datapoints in the training, validation and test sets respectively.

\begin{figure}[ht!]
    \centering
    \includegraphics[width=0.9\columnwidth]{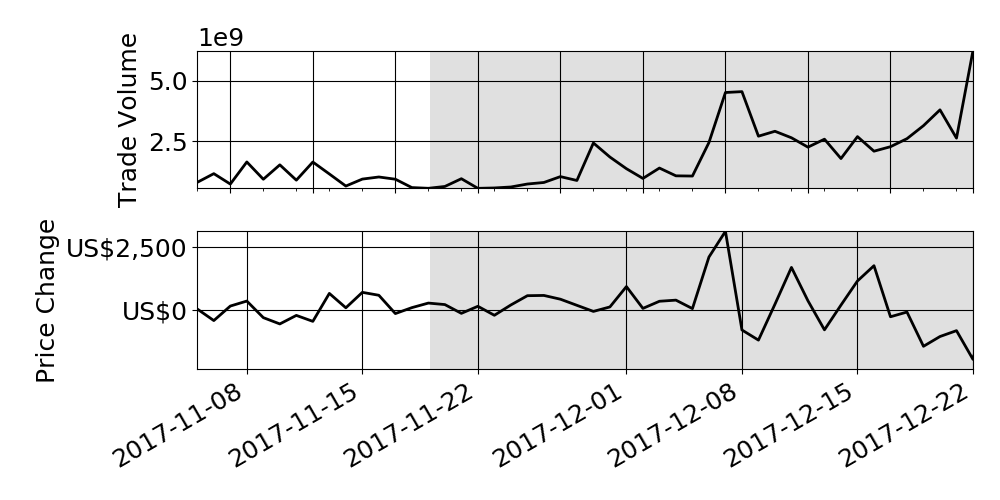}
    \caption{Plot of BTC-USD daily price change and volume of trading activity, where the shaded area being our test period 19 Nov 2017 - 22 Dec 2017}
    \label{fig:bitcoin}
\end{figure}

The test set can be split into two periods with very different market behaviours, as illustrated in Figure \ref{fig:bitcoin}. Datapoints in the test set timestamped between 19 Nov 2017 to 5 Dec 2017 have similar market behaviour to the training set. However, the succeeding period from 6 December until the end of our test period is the extremely volatile period during which the infamous 2017 Bitcoin bubble reached its height. In this paper, we will refer to the  testing period between 19 November to 5 December as \emph{Pre-Bubble} and the period from 6 Dec onwards as \emph{Bubble}. In our experiments, the models are tested on the Bubble period without any retraining on more recent data that includes the Bubble behaviour.

\subsection{Covariates \& Target Variable}

There are two types of covariates used in our model as described in the previous section: temporal covariates $\mathbf{x}_{i,:m}$ and static covariates $\mathbf{x}_i$. The temporal covariates that we chose for our experiments are a sequence of events called the \emph{order flow}, where each event is essentially an action taken by a market participant, such as placing and cancelling an order. Therefore, we define an event to be $\mathbf{x}^{(j)}_{i,\gamma}$ for all $\gamma \in \{ 1, 2, \dots m\}$, with $j \in \{ 1, 2, \dots 5\}$ as follows:

\begin{itemize}
  \item $x^{(1)}_{i,t}$ is the number of milliseconds between the arrival of $\mathbf{x}_{i,\gamma-1}$ and $\mathbf{x}_{i,\gamma}$, such that $x^{(1)}_{i,t} \geq 0$;
  \item $x^{(2)}_{i,t}$ is the size of the order, such that $x^{(2)}_{i,t}>0$;
  \item $x^{(3)}_{i,t} \in \{1,2,3\}$ is the categorical variable for the type of the order (i.e. limit price orders, open price orders, or cancellation orders);
  \item $x^{(4)}_{i,t} \in \{1,2\}$ is the categorical variable for whether it is a buy order or sell order;
  \item $x^{(5)}_{i,t}$, such that $x^{(5)}_{i,t} \geq 0$, is the price associated with the action described by the order.
\end{itemize}

The static covariates $\mathbf{x}^{(s)}_i$ are non-temporal features that are useful for predicting $y_{i,m}$. We define the static covariates used in our experiments as follows:

\begin{itemize}
  \item $x^{(s),(1)}_{i} \in \mathbb{N}$ is the hour of the day in which our target variable the price movement $y_{i,m}$ falls, which helps capture intra-day trading activity seasonality;
  \item $x^{(s),(2)}_{i} \in {1,2}$ is a categorical variable indicating which of the two currency pair (BTC-USD or BTC-EUR) datapoint $i$ belongs to.
\end{itemize}

Finally, given a sequence of orders $\mathbf{x}_{i,:m}$, our target variable for prediction is the price change $y_{i,m}$ at $\tau$ seconds after the arrival of the last order in the sequence $\mathbf{x}_{i,m}$. For our experiments, we use $\tau=15$ seconds, a good compromise between a too-small value, which could generate too many zeroes (null events), and a too-large value, which would be less accurate (predicting too far into the future).

\subsection{Implementation, Training \& Tuning}

Our proposed architecture, as described in Section \ref{section:method}, is implemented using Tensorflow. The parameters of network are fitted using the Adam optimisation algorithm with dropout on all non-recurrent connections. Within the RNNs in the architecture, we use LSTM cells \cite{hochreiter1997long}. Other RNN cells were considered but we found little difference in the output. The static covariates are appended to the temporal covariates, repeated at each sequential point. We choose the length of the temporal covariates to be $m=300$, based on our experience in this domain.

For hyperparameter tuning, we take the Bayesian approach, using a library provided by \cite{bergstra2013hyperopt}. The hyperparameters to be tuned are the number of recurrent layers, the LSTM state size, the number of dense layers in the output, the size of each output dense layers, and the embedding dimensions. The costly cross-validation procedure is implemented on a parallel GCP Kubeflow pipeline running on GPU machines to decrease the time taken by the experiments.

\subsection{Benchmark Models}

To benchmark our models we identify the following approaches in the literature as the most relevant for comparison and evaluate the performance of our models against them:

\begin{enumerate}
  \item Continuous Markov Birth-Death process \cite{cont2010stochastic}
  \item Poisson Mixture GLM \cite{jayasekare2016modeling}
\end{enumerate}

Benchmark 1 was chosen because it is a well established approach in the literature for the stochastic modelling of high-frequency price movements. In order to produce a probabilistic forecast in a form suitable for direct comparison with our proposed models, we implemented a process which draws samples from the fitted stochastic models and puts it through the emulator previously implemented for data collection to obtain an empirical distribution of the price movements. Sampling of order sizes needed for both benchmarks is done using historical order sizes.

Benchmark 2 is the only machine learning related work in the literature that produces probabilistic forecasts in the same manner as our proposed models. However, in \cite{jayasekare2016modeling}, the author used only a single covariate to predict high-frequency price movements while in our implementation we include a number of static covariates we found useful for predicting the prices through domain knowledge.

Note that since the sequential data used to train our proposed model has non-regular intervals, careful alignment of timestamps is taken to ensure the proposed and benchmark models are predicting the same target variable.

\section{Experimental Results}
\label{section:results}

The natural way to evaluate the accuracy of a probabilistic forecast would be to compute the quantile loss \cite{wen2017multi, salinas2019deepar}. However, computation of the quantile loss requires the quantile function of the random variable to be evaluated. Closed-form derivation of the quantile function for our mixture models is quite involved, and out of the scope of this paper. An alternative would be to use a Monte Carlo approach where samples are obtained from the mixture models and quantile loss is evaluated on the empirical CDF.  But it would still be difficult to evaluate the performance of the models using the quantile loss since the importance predicting the correct direction of the price movement is not accounted for. If the model assigns very low mixture probabilities to the right direction, that means we run a high risk of betting in the wrong direction (i.e. buying instead of selling when stock price is going down). Computation of the quantile loss for evaluation is therefore problematically complex, however approached.

Instead, for our experiments we propose to use a two step evaluation procedure for testing the performance of the probabilistic forecasts:

\begin{enumerate}
 \item First we evaluate the \emph{directional risk} by taking the mixture component with the highest probability to obtain directional point forecasts. The problem of evaluating the directional risk has in this way been reduced to the standard machine learning problem of comparing classification performance. We use the Matthews correlation coefficient (MCC) to evaluate the directional performance \cite{powers2011evaluation}. This metric (where a score of zero indicates random performance) is chosen because it gives a good overall assessment of the quality of the confusion matrix, and is able to handle imbalanced datasets in an effective way. The MCC does not give a misleadingly high score when a model has assigned most instances to the majority class, as would be the case for the often-used accuracy (proportion of correct class assignments) measure.

  \item Next, we separately evaluate the \emph{size risk}. For every directional point forecast that is correct, we evaluate the quantile loss for the associated distribution. In other words, we want to know: if the model gets the direction of the price movement right, how good is the subsequent prediction of the size of the movement? Since the quantiles for Poisson and Negative Binomial distributions are well-known, let $\hat{y}^\rho_{i,m}$ be the computed $\rho$ quantile for the predicted size of the price movements. Then, given the true value $y_{i,m}$ we define the $\rho$ quantile loss as

\begin{equation}
  L_{\rho} = (y_{i,m} - \hat{y}^\rho_{i,m}) (\rho \mathbb{I}_{\hat{y}^\rho_{i,m} > y_{i,m}} - (1-\rho) \mathbb{I}_{\hat{y}^\rho_{i,m} \leq y_{i,m}}  )
  \label{eq:quantile_loss}
\end{equation}

\noindent
where $\mathbb{I}_{(.)}$ is an indicator function for the given statement. For the Zero-Truncated Poisson mixture model, we only evaluate the quantile loss using Equation \ref{eq:quantile_loss} if non-zero directional price movements are correctly predicted.
\end{enumerate}

We note that although turning the probabilistic forecasts into point forecasts in Step 1 appears to defeat the purpose of the probabilistic architecture, it is a compromise we need to make in order to be able to benchmark the performance of the model using standard metrics. Later, however, in Section \ref{section:trading_scenario}, we put the models through a trading scenario to test the full probabilistic forecasts. 

\subsection{Directional Risk}

The directional risk for each model, in terms of MCC, is summarised in Table \ref{tab:compare_direction}. Poisson, Zero-Truncated Poisson (ZTP) and Negative Binomial (NB) refer to the form of mixture model used in our proposed architecture, while the benchmark models are as described in the previous section. For predicting the price change 15 seconds ahead, we can see from the table that our deep learning approach (denoted by the names of the three mixture models) outperforms the benchmark models. Comparing the different mixture models used in the output of the proposed architecture, the standard Poisson output is the least effective in modelling the directional price movements, while the ZTP output is best.

\begin{table}[ht!]
\centering
\caption{Average Matthews Correlation Coefficient of the Proposed Models and Benchmarks for $\tau=15$ in the Bubble and Pre-Bubble Test Periods}
\label{tab:compare_direction}
\begin{tabular}{@{}lccc@{}}
\toprule
Period                 & Pre-Bubble    & Bubble      \\ \midrule
Poisson                & 0.12          & 0.10           \\
Negative Binomial      & 0.14          & \textbf{0.14}          \\
Zero-Truncated Poisson & \textbf{0.16} & 0.13  \\
Benchmark 1            & 0.08          & 0.02             \\
Benchmark 2            & 0.06          & 0.01       \\ \bottomrule
\end{tabular}
\end{table}

Although we are here only comparing the performance of the directional forecast, which relies only on the mixture probabilities, one might assume that Poisson, NB and ZTP would be comparable. However, the mixture probabilities and the parameters of the distribution of each component are jointly trained through the likelihood. Hence, there is a complex interaction between the estimated component distribution parameters and the estimated mixture probabilities which for the same datapoint can lead to very different mixture probabilities being learned in individual models. We think that ZTP and NB outputs outperform Poisson here because of the explicit modelling of zero price changes in the ZTP mixture and the ability to account for overdispersion in the NB component probability distributions.

\subsection{Size Risk}

To evaluate the size risk of the models, Table \ref{tab:compare_quantile} shows the $0.5$ and $0.9$ quantile losses of the predicted probability distributions when the mixture components are correctly predicted. For easier comparison, we scale the results for the other models to a baseline model. Benchmark 2 is chosen as the baseline model for this scaling purpose as its outputs are directly comparable with our models, without the need for an emulator, as with Benchmark 1. We can see at a glance that all the models using our proposed architecture outperform the benchmark models. Between the benchmark models, although Table \ref{tab:compare_direction} shows that the directional forecast of Benchmark 1 is comparable to Benchmark 2, here we can see that Benchmark 1 struggles to forecast the size of the price movements compared to Benchmark 2.

\begin{table}[ht!]
\caption{Quantile Loss of the Proposed Models and Benchmarks in the Bubble and Pre-Bubble Test Periods, Scaled to Baseline Model}
\label{tab:compare_quantile}
\centering
\begin{tabular}{@{}lccc@{}}
\toprule
Quantile                 & \multicolumn{2}{c}{0.5} \\ \midrule
Period                 & Pre-Bubble & Bubble  \\ \midrule
Poisson     & 0.78        & 0.77   \\
Negative Binomial      & \textbf{0.65}        & \textbf{0.61}    \\
Zero-Truncated Poisson & 0.70      & 0.68     \\
Benchmark 1            & 1.23        & 1.29     \\
Benchmark 2 (Baseline)            & (1.00)        & (1.00)     \\ \midrule
Quantile                 & \multicolumn{2}{c}{0.9} \\ \midrule
Period                 & Pre-Bubble  & Bubble \\ \midrule
Poisson (Baseline)     & 0.72        & 0.70    \\
Negative Binomial      & \textbf{0.60}        & \textbf{0.56}  \\
Zero-Truncated Poisson & 0.72        & 0.71  \\
Benchmark 1            & 0.98        & 0.97  \\
Benchmark 2 (Baseline)          & (1.00)        & (1.00) \\ \bottomrule
\end{tabular}
\end{table}

Comparing the different mixture models used to obtain the outputs of our proposed architecture, we note that NB outperforms both ZTP and Poisson. This may be due to the ability of the deep NB mixture model to model the overdispersion in population of price movement sizes. Comparing the performance of each model between the Pre-Bubble and Bubble period, we can see that as the market becomes volatile, the outperformance of our models in relation to the baseline model increases. However, this increase in relative performance is only slight for the deep Poisson mixture model. The highly volatile behaviour of the market in the Bubble period may have caused a higher degree of overdispersion to arise from the clumping of the price changes. This causes the deep Poisson mixture model, which does not model the variance, to underperform compared to the deep NB model, which specifically models the variance parameter. The deep ZTP mixture model also underperforms, even though directly modelling zeroes in the mixture probabilities in this model reduces the overdispersion in the data.

Comparing the losses in the $0.5$ quantile and the $0.9$ quantile, we can see that Benchmark 1 has a lesser tendency to overpredict compared to the baseline. Overall our proposed models have a lesser tendency to overpredict compared to both benchmarks. Comparing the different proposed deep mixture models, we see that ZTP tends to overpredict more often and we think this is due to the fact that zeroes are truncated in its likelihood (i.e. distributions model $y_{i,m}>0$).

\section{Trading Simulation}
\label{section:trading_scenario}

Previously in Section \ref{section:results} we made the compromise of transforming the probabilistic directional forecast into point forecasts so that we could more directly compare our test results to those obtained from the benchmark models. In this section, we test the full probabilistic forecasts of both the direction and size of the price movements by using both the proposed models and the benchmark models in a simulated trading scenario. This section also shows how the probabilistic forecasts can be used to manage risk within automated trading, since they can be used to compute the Kelly Criterion \cite{thorp2011kelly}, a common risk management strategy.

\subsection{Trading Simulation Set-Up}

In the scenario, each model starts off with a capital of $\$10,000$. We assume that all the models have already been trained and tuned on the training and validation sets. The model are then each given a randomly sampled datapoint from the test set which is pre-sorted by timestamp. The random sampling of the datapoint is done monotonically such that, if a datapoint at time $t$ has been sampled, then datapoints timestamped before $t$ are exempted from the subsequent sampling process.

Taking this sampled datapoint as a trading opportunity, the models produce a probabilistic forecast which is then used to compute the Kelly Criterion to determine what proportion of the capital to invest in this trading opportunity such that the long-run growth of the capital is sustained. After $\tau$ seconds, we assume that the model is able to immediately exit the trade without any trading costs. We also assume that the market does not react to the trading activities of our models. After exiting the trade and updating the capital with the resulting profit or loss, we then repeat the same procedure with another sample from the test set. After $T$ iterations of this, we store the results and repeat the scenario from the beginning. We perform $K$ scenarios and observe the overall results.

The standard Kelly Criterion $f_t$ at a given point in time $t$ can be computed as follows:

\begin{equation}
f_t = s_t(\frac{\pi^2_{t+\tau}}{\hat{y}^2_{t+\tau}} - \frac{\pi^1_{t+\tau}}{\hat{y}^1_{t+\tau}})\epsilon
\label{eq:kelly_criterion}
\end{equation}

\noindent
Recall from a previous section that $\pi^1$ denotes the probability of a downward price movement and $\pi^2$ for the probability of an upward price movement, the subscript $t+\tau$ indicates the value $\tau$ seconds after current point in time, $s$ is the current price of the asset, $\hat{y_1}$ is the expected value of the price change if the price were to go down and $\hat{y_1}$ is the expected value of the price change if the price were to go up, and finally that $\epsilon$ is the risk aversion constant serving as a multiplier. If the value of $f_t$ is positive then the models take a long position (i.e. buy) in the stock. Otherwise, if it is negative, then the models take a short position (i.e. sell) the stock. We assume that the models are able to leverage risk-free loans to invest in the given trading opportunity if $|f_t|>1$.

\subsection{Results}

\begin{figure}[ht!]
    \centering
    \includegraphics[width=0.95\columnwidth]{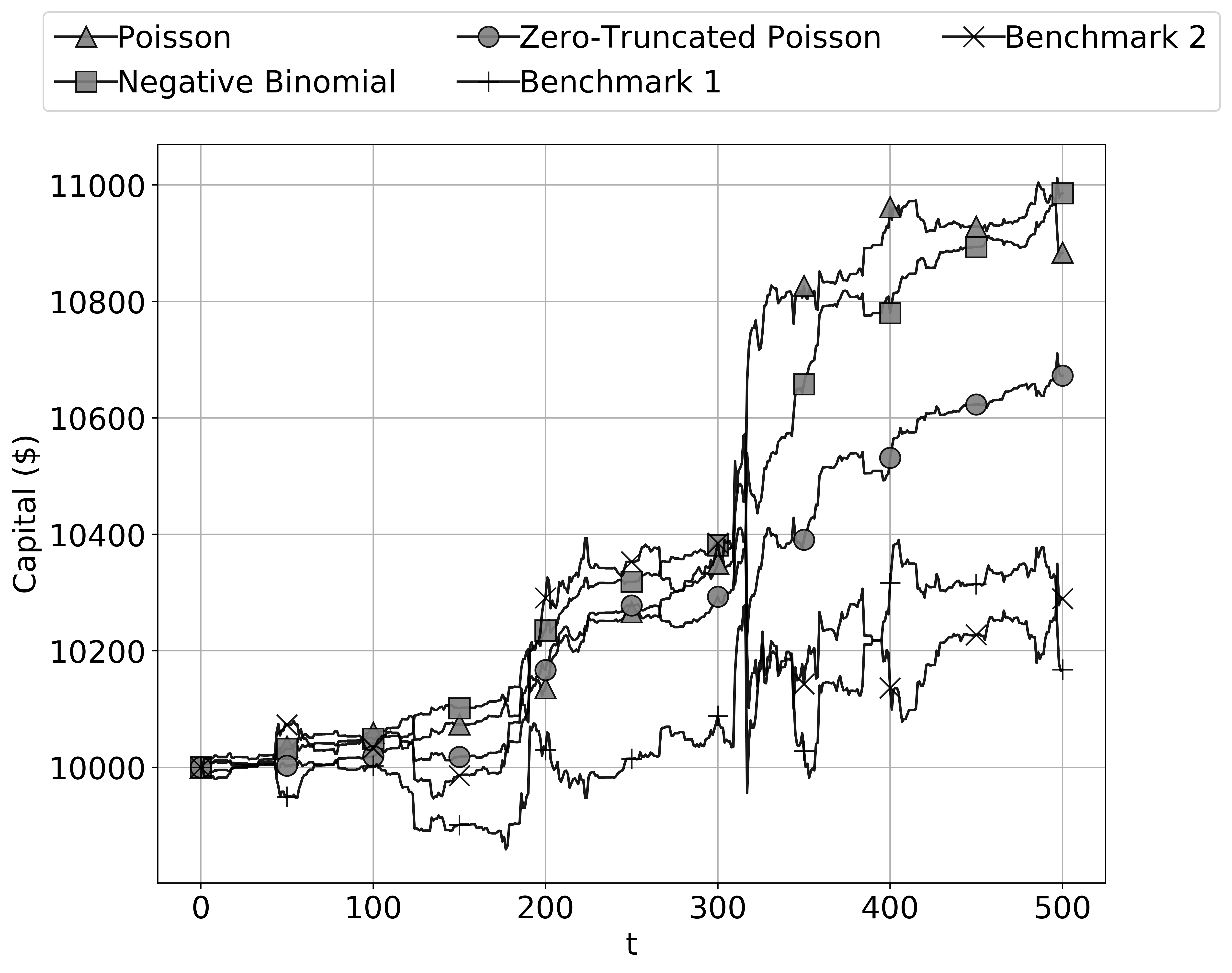}
    \caption{A single sample of the simulated trading scenario showing the change in capital due to the trading decisions made by the models}
    \label{fig:sim_one}
\end{figure}

We run the above procedure for simulating trading scenarios using the $\tau=15$ dataset for values of $T=500$ iterations in each scenario, and $K=10,000$ scenarios, with Figure \ref{fig:sim_one} showing a sample from these $K$ scenarios.

Observing the trajectories around and before the $t=320$ mark, Benchmark 2 is for this scenario performing better than the deep mixture models. However, due to an over-estimation of the Kelly Criterion caused by inaccurate probabilistic forecasts, Benchmark 2 allocates too much capital into a risky trading opportunity and ends up losing much of its capital. Around the same time, we can see that the deep ZTP mixture model also wrongly predicts the direction of the trade but does not suffer too much loss due to good probabilistic prediction of the price change when estimating the Kelly Criterion.

\begin{figure}[ht!]
    \centering
    \includegraphics[width=0.8\columnwidth]{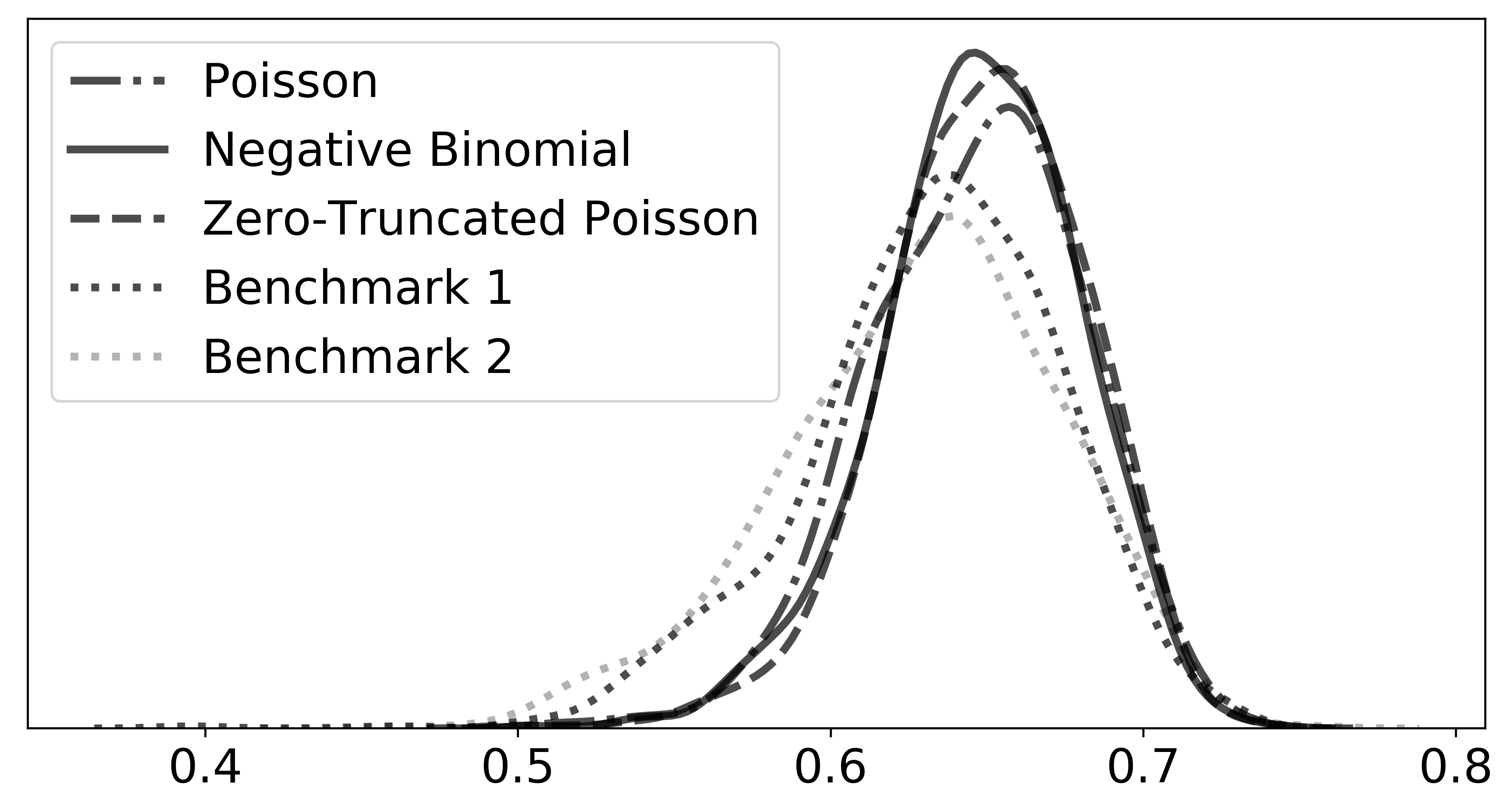}
    \caption{Empirical distribution of the final capital held by each model after 500 iterations of trading across $10,000$ trading scenarios, scaled to a hypothetical perfect prediction baseline model}
    \label{fig:sim_dist}
\end{figure}

From Figure \ref{fig:sim_one} we can observe that overall the deep models perform much better than the benchmarks in the trading simulation. However since that is only one sample from the $K$ scenarios, we cannot yet draw any conclusions. In Figure \ref{fig:sim_dist}, we show the empirical distribution of the capital held by each model at the end of each trading scenario. The values are scaled to a hypothetical baseline model that makes a perfect prediction at each iteration. At a glance we can observe that the overall deep models do better than the benchmark models. Although the mode of the benchmark model distributions is comparable to those of the deep mixture models, we can observe from the smaller peaks and right sight skew that the benchmark models are often less profitable.

We also performed paired Student t-tests on the null hypotheses that there is no difference between the profit distribution of each of the benchmark models against each of the deep mixture models. As shown in Table \ref{tab:compare_pvalue}, the null hypothesis for each test is rejected with very high confidence interval.

\begin{table}[ht!]
\centering
\caption{$p$-value of paired Student t-tests on the null hypotheses that the profit distribution for a given benchmark model is no different to the those of a deep mixture model}
\label{tab:compare_pvalue}
\begin{tabular}{@{}lcc@{}}
\toprule
                       & \multicolumn{2}{c}{p-value} \\ \midrule
                       & Benchmark 1  & Benchmark 2  \\ \midrule
Poisson                & $4.7e^{-30}$    & $8.2e^{-17}$    \\
Negative Binomial      & $9.7e^{-30}$    & $3.2e^{-16}$    \\
Zero-Truncated Poisson & $1.4e^{-39}$    & $4.0e^{-24}$      \\ \bottomrule
\end{tabular}
\end{table}

\section{Discussion}
\label{section:discussion}

We have proposed a novel architecture that allows us to produce probabilistic forecasts of the price movements of Bitcoin for use in high-frequency trading using deep mixture models. We compared the deep models, against benchmark models identified from current literature, using a proposed two-step procedure (assessing
both directional and size risk, for a trade) and also within a simulated trading scenario. In our experiments, we have shown that the proposed deep mixture models perform better in relation to both the Matthews Correlation Coefficient and the quantile loss. Also, the probabilistic forecasts produced by the deep mixture models result in statistically significantly better profit when used in a trading simulation. Note that we used the standard Kelly Criterion in the trading simulation that assumes normality; in the future we hope to test Kelly Criterion values that are derived from the different probability distributions used in the output of the proposed architecture.

Another possible direction of future work, of great interest, is the explainability of the model. Due to the increased regulatory attention given to high-frequency trading activities, black box models are being increasingly criticised. Therefore a natural extension this work would be to use black-box feature explainers such as Shapley Additive Explanations \cite{lundberg2017unified} to address the interpretation issue of the proposed models, and to understand exactly what it is that drives probabilistic price formation for the Bitcoin markets. Such a work would be valuable as it would provide an alternative data-driven view of the market microstructural behaviour of cryptocurrencies to that of existing work in quantitative finance literature. Also of interest for the future is what we can learn about the cryptocurrency market microstructure from analysis of the embeddings of categorical features of the order flow. Finally, it would also be of great interest to perform a more in-depth analysis of the suitability of different mixture models in the output of the probabilistic architecture.

\bibliographystyle{ieeetran}
\bibliography{ref}

\end{document}